\documentclass{eas}
\usepackage{graphicx}
\begin{document}

\title{Tidal Decay and circularization of the orbits of Short-Period planets}\thanks{The authors acknowledge the support of CNPq and CAPES/SECYT agreement.} 
\author{Adri\'an Rodr\'iguez}\address{Instituto de Astronomia, Geof\'isica e Ci\^encias Atmosf\'ericas, Universidade de S\~ao Paulo, Brasil}
\author{Sylvio Ferraz-Mello}\sameaddress{1}
\runningtitle{On the Orbital Decay of Short-Period planets}
\begin{abstract}

We analyze the long-term tidal evolution of a single-planet system through the use of     numerical simulations and averaged equations giving the variations of semi-major axis and eccentricity of the relative orbit. For different types of planets, we compute the variations due to the planetary and stellar tides. Then, we calculate the critical value of the eccentricity for which the stellar tide becomes dominant over the planetary tide. The timescales for orbital decay and circularization are also discussed and compared.

\end{abstract}
\maketitle
\section{Introduction}

It is well-known that tidal friction produces variations in the orbital elements of a close-in companion. In the case of an interacting pair with two extended (and tidally deformed) bodies, the tidal evolution depends on the rotational state of the bodies. We can identify two important examples: the satellite and the exoplanet cases. In the first case, the rotation of the central body (a parent planet) is generally much faster than the mean orbital motion of the satellite orbiting the primary. In the exoplanet case, the rotation of the central body (a parent star) is generally much slower than the mean orbital motion of the planet. Changes in orbital elements are also accompanied by variations in the rotation of each deformed body. The reader is referred to Ferraz-Mello \etal\ (2008) for further details.


\section{Mean orbital variation}\label{equacoes_medias}

Consider a two-body system formed by a single star and a short-period companion planet. We suppose that both bodies are able to be tidally deformed due to the mutual interaction. Tides on each body provokes variations in the elements of the relative orbit and their rotations. The equations that govern the mean orbital changes in the astrocentric orbital elements of an exoplanet are, following Ferraz-Mello \etal\ (2008) (correcting a misprint in equation (93) and neglecting the inclinations and the terms proportional to $\Omega/n$):

\begin{equation}\label{ndot}
\langle\dot{n}\rangle=\frac{9n^2k_{d\ast}m_pR_{\ast}^5|\epsilon'_{0\ast}|}{2m_{\ast}a^5}(1+23e^2+7De^2)
\end{equation}
\begin{equation}\label{edot}
\langle\dot{e}\rangle=-\frac{27nek_{d\ast}m_pR_{\ast}^5|\epsilon'_{0\ast}|}{2m_{\ast}a^5}\Bigg{(}1+\frac{7}{9}D\Bigg{)}
\end{equation}
where $n$, $a$ and $e$ are the mean orbital motion, semi-major axis and eccentricity while $m_{\ast}$,$R_{\ast},k_{d\ast}$ and $m_p$,$R_p,k_{dp}$ are the masses, radii and dynamical Love numbers of star and planet, respectively. We stress that the above equations are valid only in the case of star-exoplanet interacting pair. The main equations for the planet-satellite case are different (Ferraz-Mello \etal\ 2008). 

The parameter $D$ is defined as

\begin{equation}\label{D}
D\equiv\frac{k_{dp}}{k_{d\ast}}\frac{|\epsilon'_{2p}|}{|\epsilon'_{0\ast}|}\Bigg{(}\frac{m_{\ast}}{m_p}\Bigg{)}^2\Bigg{(}\frac{R_p}{R_{\ast}}\Bigg{)}^5
\end{equation}
where $\epsilon'_{0\ast}$ and $\epsilon'_{2p}$ are lag angles associated to the tidal waves whose frequencies are $2\Omega_{\ast}-2n$ (on the star) and $2\Omega_p-n$ (on the exoplanet), respectively. Here, $\Omega$ is the angular velocity of rotation of the tidally deformed body. The lag angles come from the delay in their response to the tide raising potentials (see Ferraz-Mello et al., 2008 for details). Several hypotheses were done to obtain the above equations. The planetary rotation was assumed in a quasi-synchronous state ($\Omega_p\sim n$) and the star rotation verifies $\Omega_{\ast}\ll n$. For the star, a \textit{linear model} is assumed in the relationship between lag angles and frequencies (Darwin 1880; Mignard 1979). In this case, each lag angle is proportional to the frequency of the corresponding tidal wave, and the coefficient of proportionality $\Delta t_{\ast}$ (time delay), is the same for all frequencies. No model needs to be assumed for the planetary tide lags, except that equal frequencies are assumed to span equal lags. The motion is supposed planar, i.e the reference and orbital planes coincide (zero obliquities). The equations are valid only up to second order in the eccentricity.

In order to simplify the notation, we define the following parameters:

\begin{eqnarray}\label{ps}
\hat{s}=\frac{9}{4}\frac{k_{d\ast}}{Q_{\ast}}\frac{m_p}{m_{\ast}}R_{\ast}^5 
\quad;
\qquad
\hat{p}=\frac{9}{2}\frac{k_{dp}}{Q_{p}}\frac{m_{\ast}}{m_p}R_p^5
\end{eqnarray}
where $Q$ are quality factors defined by $Q_{\ast}=|\epsilon_{0\ast}'|^{-1}$ and $Q_{p}=|\epsilon_{2p}'|^{-1}$ (in general $k_d$ and $Q$ are poorly known quantities for both stars and planets). Hence, 


\begin{equation}\label{D-p/s}
\frac{\hat{p}}{\hat{s}}=2\frac{k_{dp}}{k_{d\ast}}\frac{Q_{\ast}}{Q_{p}}\Bigg{(}\frac{m_{\ast}}{m_p}\Bigg{)}^2\Bigg{(}\frac{R_p}{R_{\ast}}\Bigg{)}^5=2D.
\end{equation}
Introducing (\ref{ps}) into equations (\ref{ndot}) and (\ref{edot}) and using the third Kepler law to relate $\dot{a}$ to $\dot{n}$, $\langle\dot{a}\rangle=-\frac{2a}{3n}\langle\dot{n}\rangle$, we have

\begin{equation}\label{adot}
\langle\dot{a}\rangle=-\frac{2}{3}na^{-4}[(2+46e^2)\hat{s}+7e^2\hat{p}]
\end{equation}
\begin{equation}\label{edot2}
\langle\dot{e}\rangle=-\frac{1}{3}nea^{-5}(18\hat{s}+7\hat{p}).
\end{equation}
Note that, for each equation, the terms proportional to $\hat{s}$ and $\hat{p}$ are the contribution to the total variation of the tides raised on the star and planet, respectively.

\section{Timescales of tidal evolution}\label{timescales}
It is clear from equations (\ref{adot}) and (\ref{edot2}) that the effect on the orbital elements due to the tidal interaction is to reduce the size and eccentricity of the relative orbit. In this section, we discuss the decreasing timescales of semi-major axis and eccentricity.

\subsection{Semi-major axis}\label{taua}

The decreasing timescale of the semi-major axis can be defined as $\tau_a\equiv a/|\dot{a}|$. Using equation (\ref{adot}),


\begin{equation}\label{taua_eq}
\tau_a=\frac{3}{2}n^{-1}a^{5}[(2+46e^2)\hat{s}+7e^2\hat{p}]^{-1}.
\end{equation}
If we need to know the timescale only due to planetary tides, it is enough to put $\hat{s}=0$ (or $Q_{\ast}^{-1}=0$) in the above equation to obtain

\begin{equation}\label{taua_pla}
\tau_{a}^p=\frac{3n^{-1}a^5}{14e^2\hat{p}}.
\end{equation}
Note that $\lim{\tau_{a}^p}=\infty$ as $e\rightarrow 0$, indicating that, when the only tide is the planetary, the semi-major axis stops decreasing after circularization. The contribution of the stellar tide follows a similar analysis putting $\hat{p}=0$ (or $Q_{p}^{-1}=0$). The result is

\begin{equation}\label{taua_star}
\tau_{a}^{\ast}=\frac{3n^{-1}a^5}{2(2+46e^2)\hat{s}}.
\end{equation}
Note that $\lim{\tau_{a}^{\ast}}=\frac{3n^{-1}a^5}{4\hat{s}}<\infty$ as $e\rightarrow 0$. This shows that after circularization the semi-major axis continues to decrease due to the stellar tide.

\subsection{Eccentricity}\label{taue}
The timescale of orbital circularization can be defined as $\tau_e\equiv e/|\dot{e}|$. Using (\ref{edot2})


\begin{equation}\label{taue_eq}
\tau_e=\frac{3n^{-1}a^{5}}{18\hat{s}+7\hat{p}}.
\end{equation}
As in the case of the semi-major axis, we can compute the individual timescale due to each tide. For the planetary and stellar tides we have, respectively 

\begin{eqnarray}\label{taue_pla}
\tau_{e}^p=\frac{3n^{-1}a^{5}}{7\hat{p}} ;
\qquad
\tau_{e}^{\ast}=\frac{3n^{-1}a^{5}}{18\hat{s}}.
\end{eqnarray}
Note that $\tau_{e}^p$ and $\tau_{e}^{\ast}$ are independent of $e$. Moreover, we see that 

\begin{equation}\label{taue_pla/taue_star}
\frac{\tau_{e}^p}{\tau_{e}^{\ast}}=\frac{18}{7}\frac{\hat{s}}{\hat{p}}.
\end{equation}
In order to have an idea of the numerical values of the above quantities, we give a simple example. Consider a planet-star system with masses and radii equal to those of the Sun and Jupiter. We have

\begin{equation}\label{p/s-SJ}
\frac{\hat{p}}{\hat{s}}\approx50\frac{k_{dp}}{k_{d\ast}}\frac{Q_{\ast}}{Q_{p}}.
\end{equation}
In the above equation, typical values are $2.5\times10^4<Q_{p}/k_{dp}<2.5\times10^5$ (Laskar \& Correia 2004) and $2\times10^7<Q_{\ast}/k_{d\ast}<1.5\times10^9$ (Carone \& P\"atzold 2007). Hence $\frac{k_{dp}}{k_{d\ast}}\frac{Q_{\ast}}{Q_{p}}\geq1$ and $\frac{\hat{p}}{\hat{s}}\gg1$. Then from (\ref{taue_pla/taue_star}), $\tau_{e}^p\ll\tau_{e}^{\ast}$, meaning that the tidal evolution of the eccentricity is mainly determined by planetary tide. In the case of a Super-Earth ($m_p\sim5m_{\oplus}$), we have $\frac{k_{dSE}}{k_{d\odot}}\frac{Q_{\odot}}{Q_{SE}}\gg1$ because the quality factor of a rocky planet is several orders of magnitude smaller than the corresponding value for a gaseous planet. In this case, the stellar tide is completely negligible compared with the planetary tide in producing orbital circularization.

\section{Orbital decay for different type of planets}\label{orb-decay}

\subsection{Critical eccentricity for orbital decay}\label{ec-sec}

As we have done in the case of timescales for orbital circularization, we may compare the timescales for orbital decay due to planetary and stellar tides. For that sake, using equations (\ref{taua_pla}) and (\ref{taua_star}) we obtain

\begin{equation}\label{taua_pla/taua_star}
\frac{\tau_{a}^p}{\tau_{a}^{\ast}}=\frac{(2+46e^2)}{7e^2}\frac{\hat{s}}{\hat{p}}.
\end{equation}
Note that the above timescales ratio depends on the eccentricity, which is a time-dependent quantity. We can compute a critical value of $e$ for which equation (\ref{taua_pla/taua_star}) is equal to unity. It is

\begin{equation}\label{ec-eq}
e_c=\sqrt{\frac{2}{7\hat{p}/\hat{s}-46}}\quad.
\end{equation}
If $e>e_c$ then $\tau_{a}^p<\tau_{a}^{\ast}$ and the planetary tide dominates. On the contray, if $e<e_c$ then $\tau_{a}^p>\tau_{a}^{\ast}$ and the stellar tide becomes more effective to produce orbital decay. 

\begin{figure}[ht]
\begin{centering}
\includegraphics[height=0.35\textheight,angle=270]{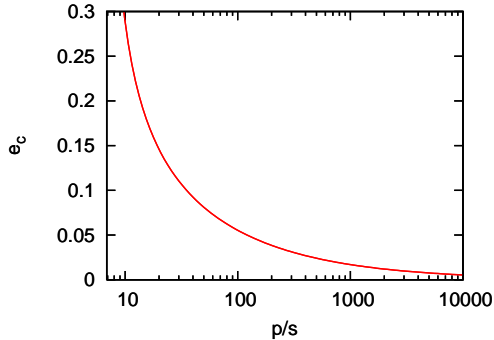}
\caption{\footnotesize A plot of equation (\ref{ec-eq}) with $\hat{p}/\hat{s}$ as independent variable. See that $\lim{e_c}=0$ as $\hat{p}/\hat{s}\rightarrow\infty$.}\label{ec-fig}
\end{centering}
\end{figure}
\noindent Figure \ref{ec-fig} is a plot of $e_c$ as a function of $\hat{p}/\hat{s}$. The minimum value of $\hat{p}/\hat{s}$ for which equation (\ref{ec-eq}) has a possible solution ($e_c<1$) is $48/7$. Note that, for high values of $\hat{p}/\hat{s}$, it is sufficient a small value of eccentricity for that the orbital decay is dominated by the planetary tide ($\tau_{a}^p<\tau_{a}^{\ast}$). On the other hand, for very small values of $\hat{p}/\hat{s}$, an eccentricity close to unity is necessary to have comparable evolutions due to planetary and stellar tides (remember that the model is valid up to second order in $e$).

\subsection{Hot Jupiters and hot Neptunes}\label{jup-nept}

In this section we analyze the tidal evolution considering two different types of close-in planets, depending on the planetary mass. In a general way we call \textit{hot Jupiter} any close-in planet for which $m_p\sim m_J$,  and \textit{hot Neptune} planets for which $m_p\sim m_N$, being $m_J$ and $m_N$ the mass of Jupiter and Neptune, respectively. From equations (\ref{taua_eq}) and (\ref{taue_eq}) we know that 

\begin{equation}\label{taua/taue}
\frac{\tau_a}{\tau_e}=\frac{1}{2}\frac{18\hat{s}+7\hat{p}}{[(2+46e^2)\hat{s}+7e^2\hat{p}]}.
\end{equation} 
It is clear from the above equation that, for small values of $e$, $\tau_a\gg\tau_e$. This indicates that the effect of tides circularize the orbit before producing any appreciable decrease in semi-major axis. Note that this fact minimizes the contribution due to the planetary tide because after circularization, only the stellar tide contributes to produce the orbital decay (see equation (\ref{adot})). Equation (\ref{ps}) shows that the stellar tide contribution is proportional to the planet mass $m_p$, where the coefficient of proportionality depends on stellar paramaters only. Hence, massive planets are more effective to have orbital decay due to the stellar tide, indicating that hot Jupiters must decay faster than hot Neptunes.

\section{Numerical simulations}\label{numerico}

In this section, we give some examples corresponding to the results obtained in the preceding sections. We adopt a linear model (Mignard 1979) to relate lags and frequencies of tidal waves for the star and the planet. It is important to mention that the Mignard's model gives the tidal force in a closed form, without any expansion in eccentricity. In that model the lag angles are given by $\epsilon_i=\nu_i\Delta t$, where $\nu_i$ is a tidal frequency and $\Delta t$ is a measure of the delay in the response of the tidal deformation, which is assumed equal for all tidal waves. It is easy to make the passage from $Q$ to $\Delta t$ provided the tidal frequency is fixed and taking into account that $Q=|\epsilon|^{-1}$. We then numerically compute the evolution of a single-planet system composed by a Sun-like star and a short-period companion planet. The selected input parameters are given in table \ref{tabla1} for two different types of planets. In each case, the values of $\hat{p}/\hat{s}$ and $e_c$ are computed through equations (\ref{D-p/s}) and (\ref{ec-eq}). For all examples showed in this section we have included a scaling factor of $10^4$ multiplying the tidal parameters of both bodies in order to accelerate the computation.

\begin{table}
\begin{center}
\begin{tabular}{|c|c|c|c|c|c|c|}
\hline
  Planet & $m_p (m_J)$  & $Q_{p}$ & $a_0$ (AU)& $e_0$ & $\hat{p}/\hat{s}$ & $e_c$\\
  \hline
  \hline
  Jupiter-like & 1.0  & $1\times10^5$ & 0.04 & 0.1 & 125 & 0.05\\
  \hline
  Neptune-like & 0.1 & $1\times10^4$ & 0.04 & 0.1 & 2.40$\times10^3$ & 0.01\\
\hline
\end{tabular}
\caption{The data for numerical simulations. Note that the value of $a_0$ corresponds to an initial orbital period of $P_0=2.92$ days. The radii values are computed imposing that $\overline{\rho}_p=\overline{\rho}_{J,N}$. For the star $m_\ast=m_{\odot}$, $R_{\ast}=R_{\odot}$ and $Q_{\ast}=1\times10^6$.}\label{tabla1}
\end{center}
\end{table}

\begin{figure}[ht]
\begin{centering}
\includegraphics[height=0.55\textheight,angle=270]{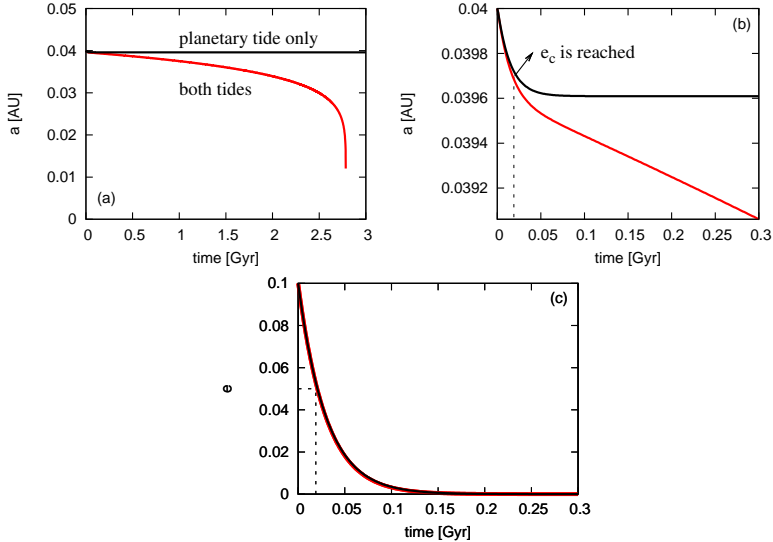}
\caption{\footnotesize (a) Long-term tidal evolution of semi-major axis for a Sun-Jupiter tidal interacting system. The cases with and without the stellar tide are showed in order to illustrate the difference between them. (b) Zoom of figure (a) showing the moment when the planetary tide becomes smaller than the stellar tide because the orbit reaches the critical eccentricity. (c) Eccentricity tidal evolution. The curves corresponding to the planetary and total tide are superimposed (see text for discussion).} 
\label{ae-1J}
\end{centering}
\end{figure}
\begin{figure}[ht]
\begin{centering}
\includegraphics[height=0.55\textheight,angle=270]{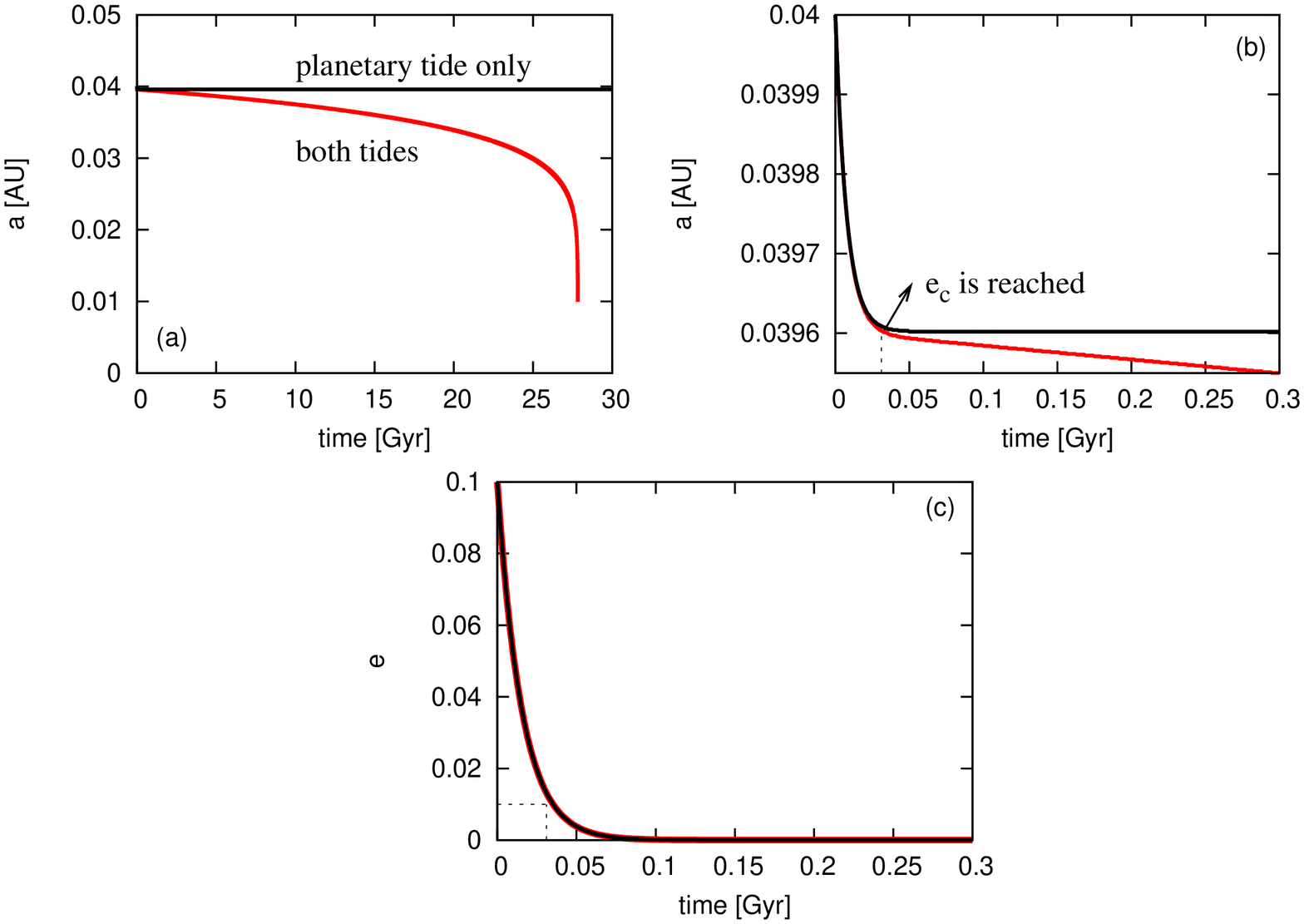}
\caption{\footnotesize (a) Neptune-like planet semi-major axis tidal evolution. See that the small planetary mass obliges to a much slower semi-major axis decay compared with the Jupiter case. It is important to note that $\tau_a$ is higher than the age of the universe. (b) Zoom of figure (a).(c) Eccentricity tidal evolution. Note the moment for which $e=e_c=0.01$.}\label{ae-5T}
\end{centering}
\end{figure}
Figure \ref{ae-1J} shows the long-term tidal evolution of semi-major axis and eccentricity in the case of a hot Jupiter around a Sun-like star. One may note that the timescale for orbital circularization $\tau_e$ is much smaller than the timescale for orbital decay $\tau_a$, as discussed in section \ref{jup-nept}. The figure also shows the comparison of the cases with and without stellar tide. It is well appreciable how the inclusion of the stellar tide changes the evolution of the semi-major axis, compared with the case where only the planetary tide is considered. Figure \ref{ae-1J}b is a zoom of \ref{ae-1J}a and it shows the time from which the stellar tide becomes dominant. Comparing to figure \ref{ae-1J}c, we note that this time corresponds to the moment in which $e\sim e_c=0.05$. When $e>0.05$, the planetary tide dominates. However, since $\tau_e^p\ll\tau_a^p$, the orbit reaches very quickly the critical eccentricity and after that the stellar tide starts to be dominant. We note that the eccentricity tidal evolution is almost independent on the inclusion of the stellar tide. This happens because $\tau_e^{p}\ll\tau_e^{\ast}$ and therefore the contribution of the stellar tide to the orbital circularization is almost negligible. 

Figure \ref{ae-5T} shows the long-term tidal evolution for a Neptune-like planet interacting with the same star under the same initial conditions ($m_p\sim2m_N$). As in the case of a Jupiter-like planet, the semi-major axis evolution is different when the stellar tide is added to the planetary tide. However, in this case, the difference is not so apparent as in the case of a Jupiter-like planet, at least for the same timescale (compare figure \ref{ae-5T}a with \ref{ae-1J}a). Note that $\tau_{aN}\sim10 \tau_{aJ}$. On the other hand $e_{cN}=\frac{1}{5}e_{cJ}$, meaning that the planetary tide dominates almost until total circularization. After that point, the stellar tide controls the orbital decay. Moreover, since $2m_N\ll m_J$, the influence of the stellar tide in the case of Neptune-like planets is not so effective in producing strong damping in semi-major axis as in the case of a Jupiter-like planet. This was already anticipated in section \ref{jup-nept} where we have seen that $\hat{s}\propto m_p$. The eccentricity evolution does not show any appreciable change when the stellar tide is included, as we see in figure \ref{ae-5T}b. This is because the relation $\tau_e^{p}\ll\tau_e^{\ast}$ is more closely satisfied for the case of Neptune-like planets (see section \ref{taue}). Note also that $\tau_{eN}<\tau_{eJ}$.


\section{Conclusion}\label{conclu}

The main conclusion of this communication is that, in terms of timescales, the stellar tide plays an important role to determine the orbital decay of hot Jupiter planets. In the case of hot Neptune-like planets the contribution of the stellar tide is smaller but still important. The timescale for orbital decay is not sensitive to the value of $Q_{p}$, since the long-term tidal evolution of semi-major axis is controlled by the stellar tide. The orbital circularization is well-determined only by the planetary tide and is $Q_{p}$ dependent. The computation of $e_c$ enable us to known the moment for which the stellar tide becomes dominant.



\end{document}